\documentclass[a4paper,fleqn,usenatbib]{mnras}

\usepackage{newtxtext,newtxmath}

\usepackage[T1]{fontenc}
\usepackage{ae,aecompl}

\usepackage{booktabs,caption,fixltx2e}
\usepackage[flushleft]{threeparttable}

\usepackage{graphicx}	
\usepackage{amsmath}	
\usepackage{amssymb}	

\title[Missing dust signature in the CMB]{Missing dust signature in the cosmic microwave background}

\author[V. Vavry\v{c}uk]{
V\'{a}clav Vavry\v{c}uk,$^{1}$\thanks{E-mail: vv@ig.cas.cz}
\\
$^{1}$Institute of Geophysics, The Czech Academy of Sciences, Bo\v{c}n\'{i} II, Praha 4, 14100, Czech Republic\\
}

\date{Accepted May 4, 2017. Received April 29, 2017; in original form March 5, 2017}

\pubyear{2017}

\begin{document}
\label{firstpage}
\pagerange{\pageref{firstpage}--\pageref{lastpage}}
\maketitle
\begin{abstract}
I examine a possible spectral distortion of the Cosmic Microwave Background (CMB) due to its absorption by galactic and intergalactic dust. I show that even subtle intergalactic opacity of  $1 \times 10^{-7}\, \mathrm{mag}\, h\, \mathrm{Gpc}^{-1}$ at the CMB wavelengths in the local Universe causes non-negligible CMB absorption and decline of the CMB intensity because the opacity steeply increases with redshift. The CMB should be distorted even during the epoch of the Universe defined by redshifts $z < 10$. For this epoch, the maximum spectral distortion of the CMB is at least $20 \times 10^{-22} \,\mathrm{Wm}^{-2}\, \mathrm{Hz}^{-1}\, \mathrm{sr}^{-1}$ at 300 GHz being well above the sensitivity of the COBE/FIRAS, WMAP or Planck flux measurements. If dust mass is considered to be redshift dependent with noticeable dust abundance at redshifts 2-4, the predicted CMB distortion is even higher. The CMB would be distorted also in a perfectly transparent universe due to dust in galaxies but this effect is lower by one order than that due to intergalactic opacity. The fact that the distortion of the CMB by dust is not observed is intriguing and questions either opacity and extinction law measurements or validity of the current model of the Universe. 
\end{abstract}

\begin{keywords}
cosmic background radiation -- dust, extinction -- early Universe -- galaxies: high redshift -- galaxies: ISM -- intergalactic medium 
\end{keywords}%

\section{Introduction}
Observations of the Cosmic Microwave Background (CMB) based on rocket measurements of \cite{Gush1990} and  FIRAS on the COBE satellite \citep{Mather1990, Fixsen1996} proved that the CMB has almost a perfect thermal black-body spectrum with an average temperature of $T = 2.728 \pm 0.004\, \mathrm{K}$ \citep{Fixsen1996}. The accuracy was improved using the WMAP data, which yielded an average temperature of $T = 2.72548 \pm 0.00057 \, \mathrm{K}$ \citep{Fixsen2009}.
Observed tiny large-scale variations of the CMB temperature of $\pm0.00335 \,\mathrm{K}$ are attributed to the motion (including rotation) of the Milky Way relative to the Universe \citep{Kogut1993}. The small-scale variations of $\pm 300 \,\mu\mathrm{K}$ traced, for example, by the WMAP \citep{Bennett2003,Hinshaw2009, Bennett2013}, ACBAR \citep{Reichardt2009}, and BOOMERANG \citep{MacTavish2006} instruments using angular multipole moments are attributed to basic properties of the Universe as its curvature or the dark-matter density \citep{Spergel2007, Komatsu2011}.

Since the CMB as a relic radiation of the Big Bang experienced different epochs of the Universe, it interacted with matter of varying physical and chemical properties. Distortions of the CMB due to this interaction comprise the $\mu$-type (at $z \gtrsim \, 10^5$) and $y$-type (at $z \lesssim 10^4$) distortions related to the photon-electron interactions, distortions produced by the reionized IGM, and presence of galactic and extragalactic foregrounds \citep{Wright1981, Chluba2012, DeZotti2016}. The foreground contamination of the CMB due to diffuse emission of intergalactic dust thermalized by absorption of starlight was estimated, for example, by \citet{Imara2016}. They found that the predicted contamination is under the detection of the COBE/FIRAS experiments \citep{Mather1994, Fixsen1996} but it should be recognized in observations of the Primordial Inflation Explorer (PIXIE; \citeauthor{Kogut2014} \citeyear{Kogut2014}) and the Polarized Radiation Imaging and Spectroscopy Mission (PRISM; \citeauthor{Andre2014} \citeyear{Andre2014}) that would exceed the spectral sensitivity limits of COBE/FIRAS by 3-4 orders of magnitude.

Another possible origin of distortion of the CMB related to galactic and intergalactic dust is absorption of the CMB by dust. Absorbing properties of dust grains have been discussed by \citet{Wright1987, Wright1991, Henning1995, Stognienko1995} and others, who pointed out that the long-wavelength absorption of needle-shaped conducting grains or complex fractal or fluffy dust aggregates might provide a sufficient opacity for the CMB. Hence, it is worth to model the CMB attenuation by dust and to check if is detectable or not. In this paper, I study the spectral and total distortions of the CMB due to absorption by dust. I find that the imprint of cosmic dust in the CMB predicted by theory is not negligible; however, it is missing in observations even though it is above their current detection level.  

\section{Theory}
%
\subsection{Optical depth}
Effective optical depth $\tau(z)$ for light emitted at redshift $z$ is expressed as \citet[his equation 13.42]{Peebles1993}
\begin{equation}\label{eq1}
\tau\left(z\right) = \int_0^{z}n_D \sigma \left(1+z'\right)^2  
\,\frac{c}{H_0} \frac{dz'}{E\left(z'\right)} \,,
\end{equation}
where $n_D$ is the comoving dust number density, $\sigma$ is the attenuation cross-section, $E\left(z\right)$ is the dimensionless Hubble parameter  
\begin{equation}\label{eq2}
E\left(z\right) = \sqrt{\left(1+z\right)^2\left(1+\Omega_m z\right)-z\left(2+z\right)\Omega_{\Lambda}} \,\, ,
\end{equation}
$c$ is the speed of light, $H_0$ is the Hubble constant, $\Omega_m$ is the total matter density, $\Omega_{\Lambda}$ is the dimensionless cosmological constant. 

Eq. (1) can be rewritten using galactic and intergalactic attenuation coefficients $\varepsilon^G$ and $\varepsilon^{IG}$ as 
\begin{equation}\label{eq3}
\tau\left(z\right) = \frac{c}{H_0}\int_0^{z}\left(\varepsilon^G + \varepsilon^{IG}\right) \left(1+z'\right)^2  
\, \frac{dz'}{E\left(z'\right)} \,,
\end{equation}
where 
\begin{equation}\label{eq4}
\varepsilon^G = \frac{\kappa}{\gamma} \,,
\end{equation}
$\kappa$ is the mean galactic opacity, $\gamma$ is the mean free path of a light ray between galaxies in the comoving space
\begin{equation}\label{eq5}
\gamma\left(z\right) = \frac{1}{n \pi a^2} \, \, ,
\end{equation}
$a$ is the mean galaxy radius, and $n$ is the galaxy number density in the comoving space. Eq. (3) is valid for frequency-independent attenuation. Considering the '$\lambda^{-\beta}$ extinction law', where $\lambda$ is the wavelength of light \citep{Mathis1990,Calzetti1994, Charlot2000, Draine2003}, we can express the galactic and intergalactic attenuations at frequency $\nu$ using the reference quantities related to observed frequency $\nu_0$,
\begin{equation}\label{eq6}
\varepsilon_\nu^G = \nu^{\beta}\, \varepsilon_0^G \,\, ,\, \, 
\varepsilon_\nu^{IG} = \nu^{\beta}\, \varepsilon_0^{IG}\, \, . 
\end{equation}
Eq. (3) is then modified to
\begin{equation}\label{eq7}
\tau_\nu\left(z\right) = \frac{c}{H_0}
\left(\frac{\nu}{\nu_0}\right)^{\beta}\int_0^{z} 
\left(\varepsilon_0^G+\varepsilon_0^{IG}\right)
\left(1+z'\right)^{2+\beta}  
\, \frac{dz'}{E\left(z'\right)} \, ,
\end{equation}
expressing the fact that light is more attenuated at high $z$ because of its shift to high frequencies.

\subsection{Extinction of the CMB} 
Assuming the CMB to be a perfect blackbody radiation, its spectral intensity (i.e., energy received per unit area from a unit solid angle in the frequency interval $\nu$ to $\nu + d\nu$, in $\mathrm{W m}^{-2}\,\mathrm{Hz}^{-1} \mathrm{sr}^{-1}$) is described by the Planck's law
\begin{equation}\label{eq8}
I_\nu = \frac{2h\nu^3}{c^2} \frac{1}{e^{h\nu/k_B T_\mathrm{CMB}}-1} \,,
\end{equation}
where $\nu$ is the frequency, $T_\mathrm{CMB}$ is the CMB temperature, $h$ is the Planck constant, $c$ is the speed of light, and $k_B$ is the Boltzmann constant. Since the CMB is attenuated by galactic and intergalactic opacity, we can evaluate the distortion of the spectral CMB intensity at frequency $\nu$ along light ray coming from redshift $z$ as
\begin{equation}\label{eq9}
\Delta I_\nu\left(z\right) = I_\nu\,\left(1-e^{-\tau_\nu\left(z\right)} \right) \, ,
\end{equation}
where $\tau_\nu$ and $I_\nu$ are defined in eqs (7) and (8). Consequently, the reduction of the total CMB intensity ($\mathrm{in \, W m}^{-2}\,\mathrm{sr}^{-1}$) is
\begin{equation}\label{eq10}
\Delta I\left(z\right) = \int \Delta I_\nu d\nu \, .
\end{equation}

Evaluating eqs (9) and (10) for different redshifts $z$, we can predict the distortion of the CMB intensity by the opacity of the Universe when going back in cosmic time up to redshift $z$. Such approach is advantageous because it suppresses uncertainties in observed parameters needed in calculations. We start at present time, when the galactic and intergalactic opacities are best constrained from observations, and gradually extrapolate the prediction to higher redshifts.

\section{Opacity observations}
In order to evaluate the CMB distortion due to absorption by dust, we need estimates of the dust mass in the Universe and its history. The most straightforward way is to use observations of the galactic and intergalactic opacities at visual wavelengths mapping the distribution of dust in galaxies and intergalactic space and relate the visual and CMB opacities using the extinction law describing the dependence of attenuation of light on wavelength. 

\subsection{Galactic and intergalactic opacities }
The opacity of galaxies depends basically on their type and age (for a review, see Calzetti, 2001). The most transparent galaxies are elliptical with an effective extinction $A_V$ of $0.04 - 0.08$ mag. The light extinction by dust in spiral and irregular galaxies is higher \citep{Gonzalez1998, Holwerda2005a, Holwerda2005b, Holwerda2013,Holwerda_Keel2013}. Typical values for the inclination-averaged extinction are: $0.5 - 0.75$ mag for Sa-Sab galaxies, $0.65 - 0.95$ mag for the Sb-Scd galaxies, and $0.3 - 0.4$ mag for the irregular galaxies at the B-band \citep{Calzetti2001}. Considering the relative frequency of galaxy types in the Universe, we can average the visual extinctions of individual galaxy types and calculate the mean visual extinction and the mean visual galactic opacity. According to \citet{Vavrycuk2017}, the average value of visual opacity $\kappa_V$ is about $0.22 \pm 0.08$ at $z=0$. 

The intergalactic opacity is lower by several orders than the galactic opacity being observed, particularly, in galaxy halos and in cluster centres \citep{Menard2010a}. The opacity in the galaxy clusters has been measured by reddening of background objects behind the clusters \citep{Chelouche2007,Bovy2008, Muller2008}. The intergalactic opacity can also be measured by correlations between the positions of low-redshift galaxies and high-redshift QSOs. \citet{Menard2010a} correlated the brightness of $\sim85.000$ quasars at $z > 1$ with the position of 24 million galaxies at $z \sim 0.3$ derived from the SDSS. The estimated value of $A_V$ is about 0.03 mag at $z = 0.5$ and about $0.05 - 0.09$ mag at $z = 1$. A consistent opacity is reported by \citet{Xie2015} who investigated the redshifts and luminosity of the quasar continuum of $\sim 90.000$ objects. The authors estimated the visual opacity to be $\sim 0.02 \,h \, \mathrm{Gpc}^{-1}$ at $z < 1.5$. As mentioned by \citet{Menard2010b} such opacity is not negligible and can lead to bias in determining cosmological parameters if ignored.

\subsection {Evolution of opacity with redshift}

The galactic and intergalactic opacities depend on redshift. First, they increase with redshift due to the expansion of the Universe. This geometrical effect has already been taken into account in eq. (1) by considering an increasing dust density with redshift because the Universe occupied a smaller volume in its early epoch. Second, a redshift-dependent formation and evolution of global dust mass in galaxies and in intergalactic space must be taken into account. 

Observations indicate that interstellar dust mass $M_d$ is strongly linked to the star formation rate (SFR) of galaxies. \citet{Cunha2010} analysed 3258 low-redshift SDSS galaxies with $z<0.2$ and reported the relation $M_d \sim \mathrm{SFR}^{1.1}$. \citet{Calura2017} extended the dataset with high-redshift galaxies from \citet{Santini2010} and found a similar relation with a slightly lower slope of $\sim$ 0.9. The same slope is reported also by \citet{Hjorth2014}. Adopting the $M_d-\mathrm{SFR}$ relation, we deduce from the SFR history (see Fig. 1) that the global dust mass steeply increases for $z < 2-2.5$, it culminates at $z=3-4$ and then it starts to decline \citep{Madau1996, Hopkins2006, Madau2014, Popping2016}. The decline is not, however, substantially steep because dust is reported even in star-forming galaxies at redshifts of $z>5$ \citep{Casey2014}. Based on observations of the Atacama Large Millimeter Array (ALMA), \citet{Watson2015} investigated a galaxy at $z>7$ highly evolved with a large stellar mass and heavily enriched in dust. Similarly, \citet{Laporte2017} analysed a galaxy at a photometric redshift of $z \sim 8$ with a stellar mass of $\sim 2\times 10^9 M_{\sun}$, a SFR of $\sim 20 \, M_{\sun}\,\mathrm{yr}^{-1}$ and a dust mass of $\sim 6\times10^6 M_{\sun}$.

\subsection {Extinction law}
The light extinction due to absorption by dust is frequency dependent (see Fig. 2). In general, it decreases with increasing wavelength but displays irregularities. The extinction curve for dust in the Milky Way can be approximated for infrared wavelengths between $\sim 0.9 \mu$m and $\sim 5 \mu$m by a power-law $A_\lambda \sim \lambda^{-\beta}$ with $\beta$ ranging between 1.61 and 1.81 \citep{Draine2003, Draine2011}. At wavelengths of 9.7 and 18 $\mu$m, the absorption displays two distinct maxima attributed to silicates \citep{Mathis1990, Li2001, Draine2003}. At longer wavelengths, the extinction curve is smooth obeying a power-law with $\beta = 2$. This decay is also predicted by the Mie theory modelling graphite or silicate dust grains as small spheres or spheroids with sizes up to 1 $\mu$m \citep{Draine1984}. However, \citet{Wright1982, Henning1995, Stognienko1995} and others point out that the long-wavelength absorption also depends on the shape of the dust grains and that needle-shaped conducting grains or complex fractal or fluffy dust aggregates can provide higher long-wavelength opacity with the power-law described by $0.6 < \beta < 1.4$ \citep{Wright1987}. 

\begin{figure}
\centering
\includegraphics[angle=0,width = 8 cm,trim = {140 130 120 130},clip]{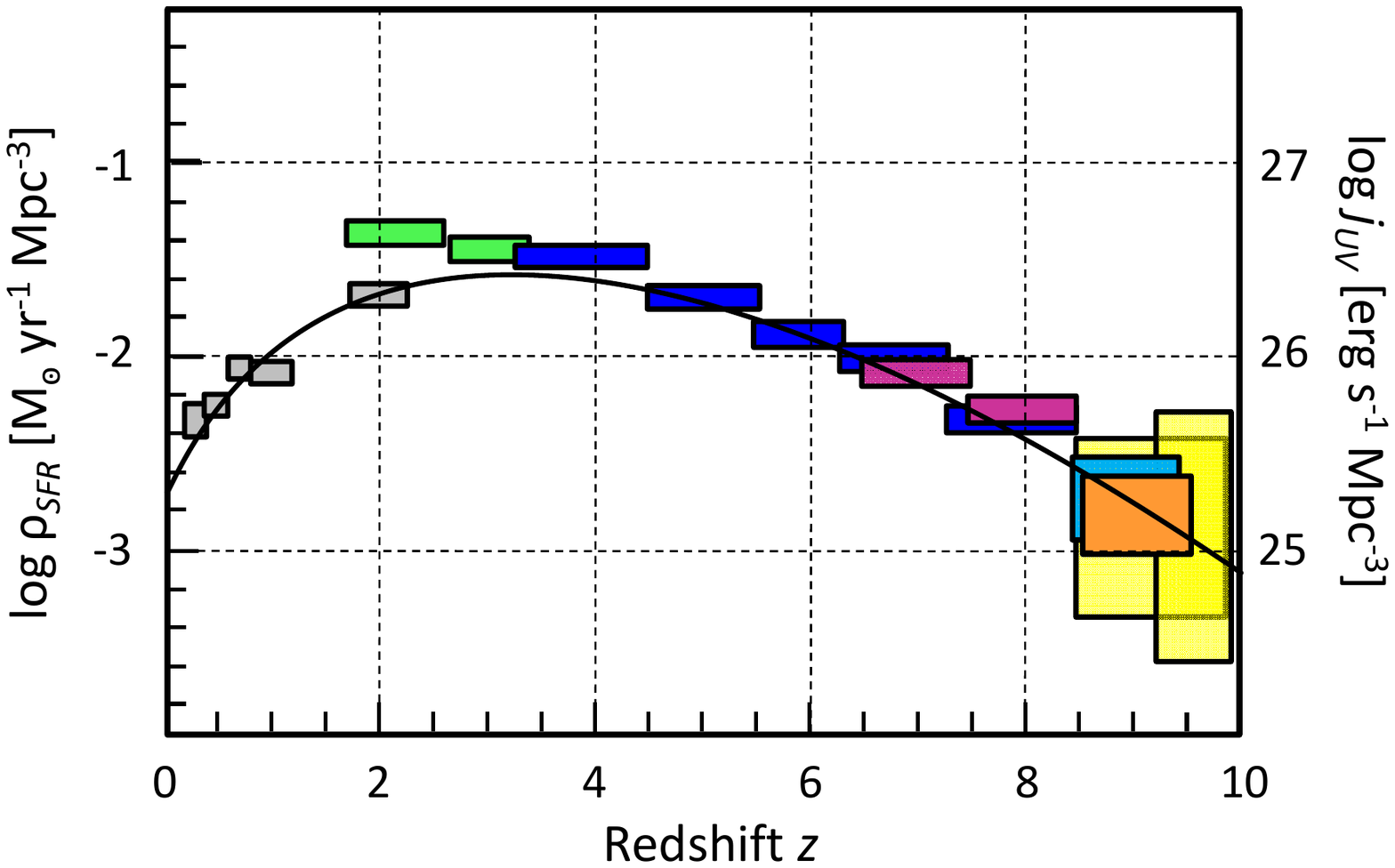}
\caption{
The SFR traced by the UV luminosity density as a function of redshift. Observations are taken from \protect \cite{Schiminovich2005} (grey rectangles), 
\protect \cite{Reddy2009} (green rectangles), \protect \cite{Bouwens2014a} (blue rectangles), \protect \cite{McLure2013} (magenta rectangles), \protect \cite{Ellis2013} (orange rectangle), \protect \cite{Oesch2014} (light blue rectangle), and \protect \cite{Bouwens2014b} (yellow rectangles). The solid black line reproduces the SFR evolution used in the modelling of evolution of dust mass.}
\label{fig:1}
\end{figure}
%

\begin{figure}
\includegraphics[angle=0,width= 7.5 cm,trim=50 50 100 80]{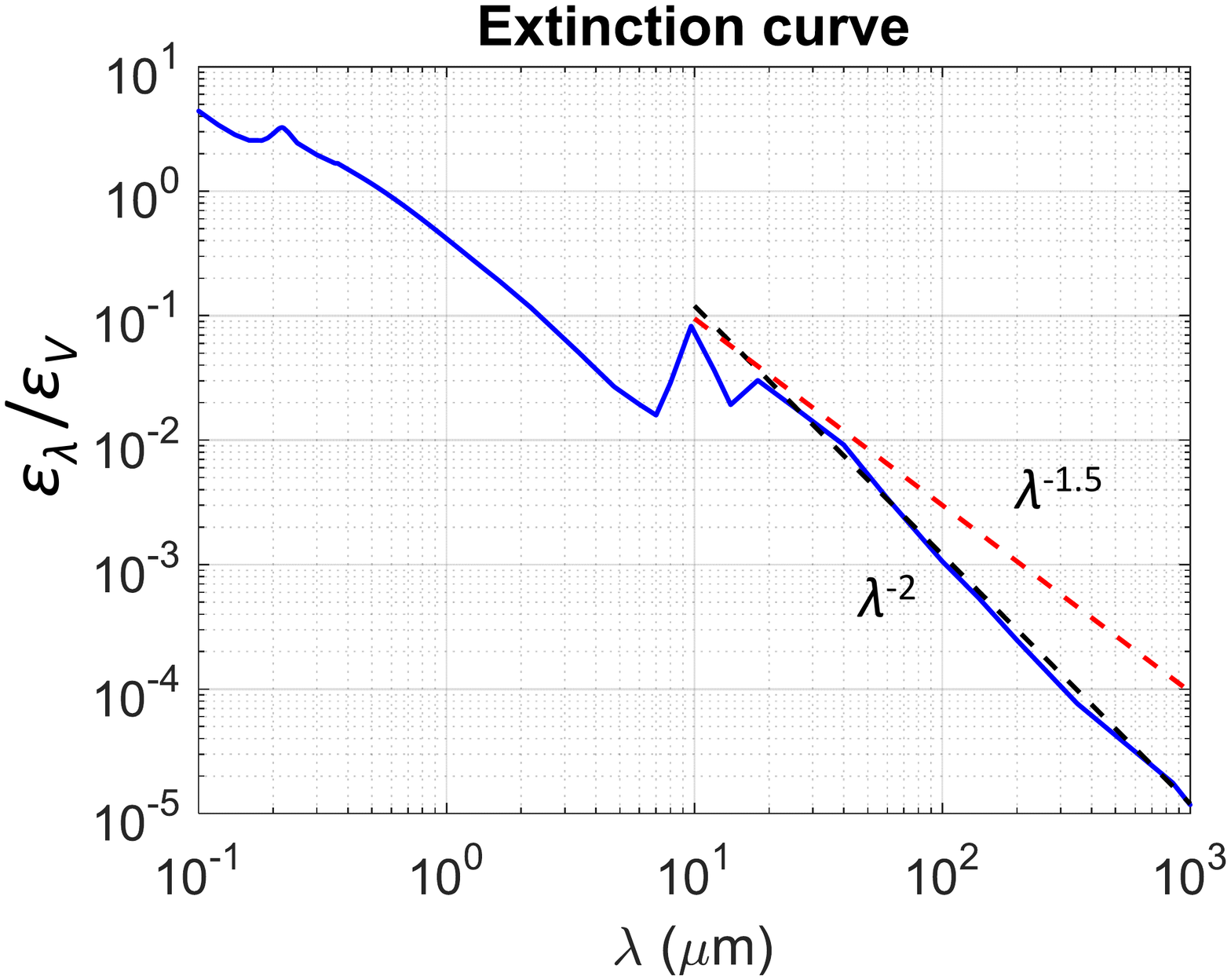}
%
\caption{
Normalized frequency-dependent attenuation \citep[his Tables 4-6]{Draine2003}. The black and red dashed lines show the long-wavelength asymptotic behaviour predicted by the power law with $\beta = 2$ and $\beta = 1.5$.
}
\label{fig:2}
\end{figure}

\begin{table*}
%
%
\caption{Input parameters for modelling}  
\label{Table:1}      
\centering
\begin{tabular}{c c c c c c c c c c}  
%
%
\hline\hline                 
%
$a$ & $n$ & $\gamma$ & $\kappa_V$ & $\beta$ & $\varepsilon_V^G$ &
$\varepsilon_V^{IG}$ & $\varepsilon_\mathrm{CMB}/\varepsilon_V$ &  $\varepsilon_\mathrm{CMB}^G$ &  $\varepsilon_\mathrm{CMB}^{IG}$
\\
%
%
 (kpc) & $(h^3\,\mathrm{Mpc}^{-3})$ & $(h\,\mathrm{Gpc}^{-1})$ &  & & $(h\,\mathrm{Gpc}^{-1})$ & $(h\,\mathrm{Gpc}^{-1})$ &  & $(h\,\mathrm{Gpc}^{-1})$ & $(h\,\mathrm{Gpc}^{-1})$ \\    
%
%
\hline                        
%
10 & 0.02 & 160 & 0.22 & 2.0 & $1.4 \times 10^{-3}$ &  $9.2 \times 10^{-3}$ & $1.0 \times 10^{-5}$ & $1.4 \times 10^{-8}$ & $9.2 \times 10^{-8}$ \\
%
%
\hline                                  
\end{tabular}
%
%
\begin{tablenotes}
\item 
$a$ is the mean effective radius of galaxies, 
$n$ is the comoving number density of galaxies, 
$\gamma$ is the mean free path between galaxies,
$\kappa_V$ is the mean visual opacity of galaxies, 
$\beta$ is the slope in the extinction law,
$\varepsilon_V^G$  is the visual galactic attenuation coefficient defined in eq. (4),
$\varepsilon_V^{IG}$ is the visual intergalactic attenuation coefficient,
$\varepsilon_\mathrm{CMB}^G$ and $\varepsilon_\mathrm{CMB}^{IG}$ are the galactic and intergalactic attenuation coefficients at the CMB wavelengths.
\end{tablenotes}

\end{table*}

\begin{figure*}
\includegraphics[angle=0,width=15cm, trim= 60 130 60 140, clip = true]{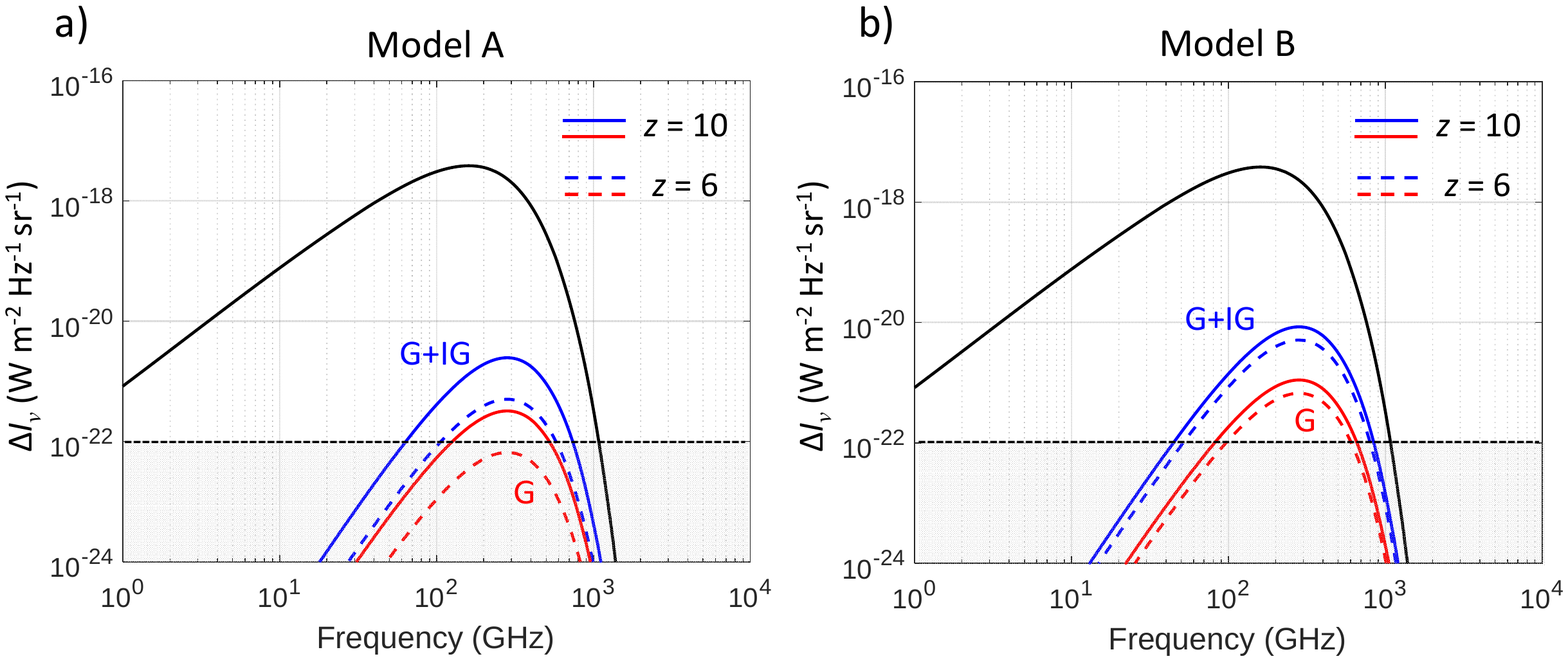}
\caption{
The spectral CMB distortion for Model A (a) and Model B (b). The full black line shows the spectral CMB intensity. Full blue/red lines - $z_\mathrm{max} = 10$, dashed blue/red lines - $z_\mathrm{max} = 6$. Blue lines - distortions due to galactic and intergalactic dust (G+IG), red lines - distortions due galactic dust (G). The grey area marks intensities which are under the sensitivity of the COBE/FIRAS measurements at 300 GHz \citep{Fixsen1996}. Cosmological parameters:  $H_0 = 70 \,\mathrm{km}\,\mathrm{s}^{-1} \mathrm{Mpc}^{-1}$, $\Omega_m = 0.3$, and $\Omega_\Lambda = 0.7$.
}
\label{fig:3}
\end{figure*}

\begin{figure}
\includegraphics[angle=0,width=7.8 cm,trim=70 60 120 80, clip = true]{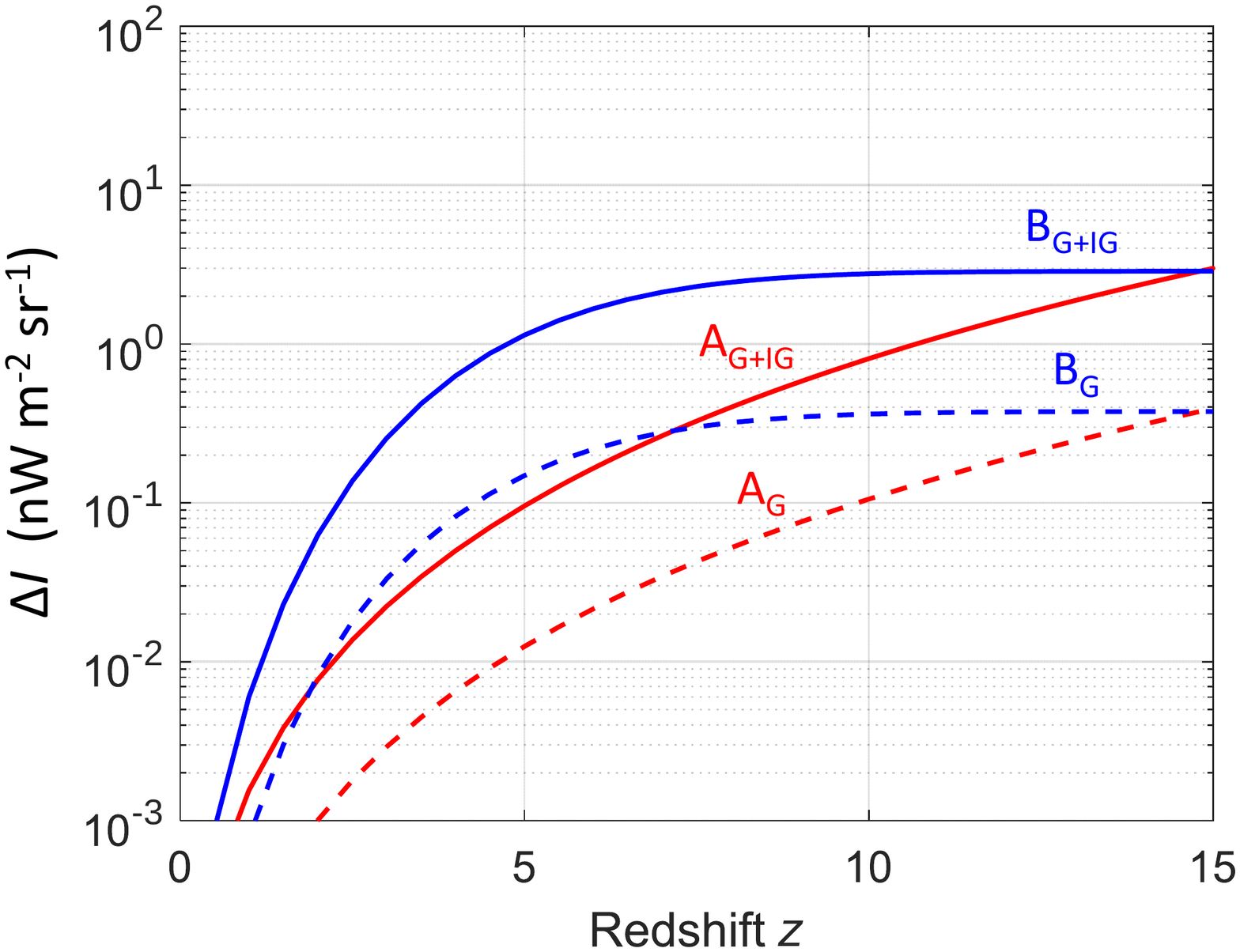}
%
\caption{
The total CMB distortion as a function of redshift for Model A (red) and Model B (blue). Full lines - distortions due to galactic and intergalactic dust (G+IG), dashed lines - distortions due galactic dust (G).
}
\label{fig:4}
\end{figure}

\section {Predicted CMB distortion}
I consider the intergalactic opacity at visual wavelengths of 0.01 $\mathrm{mag}\, h\, \mathrm{Gpc}^{-1}$ that is a twice lower value than that reported by \citet{Xie2015}. The ratio of the CMB and visual attenuation $\varepsilon_{\mathrm{CMB}}/\varepsilon_V$ of $1 \times 10^{-5}$ is taken from \citet{Mathis1990} and \citet{Draine2003}. Actually, this ratio is very low being obtained for a steep decrease of attenuation at long wavelengths ($\beta = 2$). Realistic values for dust particles with complex shapes might be higher by one order \citep[$\beta = 1.5$]{Wright1987}. I intentionally use the low value of $\varepsilon_{\mathrm{CMB}}$ in order to be sure that the predicted level of the CMB distortion is the lower threshold of expected values. 

The CMB distortion is calculated for two models. Model A is based on an assumption that the comoving dust density is independent of redshift. Model B adopts an interstellar and intergalactic dust density evolving with redshift in accordance with the SFR (see Fig. 1). The spectral and total CMB distortions are calculated using eqs (9) and (10) with parameters summarized in Table 1. In calculations, both of the galactic and intergalactic opacity (G+IG) or the galactic opacity only (G) are considered.

Fig. 3 shows the spectral CMB intensity and its corresponding distortion produced by dust in the epoch of $0 < z < z_{\mathrm{max}}$ with $z_{\mathrm{max}}$ of 6 and 10. As expected, the distortion increases with  increasing $z_{\mathrm{max}}$, but the effect of dust absorption is visible even for $z_{\mathrm{max}}$ of 6. The distortion is more pronounced for Model B than for Model A. This is caused by abundance of dust for $z \sim 2-4$ considered in Model B but neglected in Model A. The maximum distortion is observed at frequency of 300 GHz and reaches a value of $5.1 \times 10^{-22} \,\mathrm{Wm}^{-2}\, \mathrm{Hz}^{-1}\, \mathrm{sr}^{-1}$ for Model A and $51.0 \times 10^{-22} \,\mathrm{Wm}^{-2}\, \mathrm{Hz}^{-1}\, \mathrm{sr}^{-1}$ for Model B. These values exceed the detection level of the COBE/FIRAS (absolute sensitivity of $\sim 1-2\times10^{-22}\,\mathrm{Wm}^{-2}\, \mathrm{Hz}^{-1}\, \mathrm{sr}^{-1}$, \citeauthor{Fixsen1996} \citeyear{Fixsen1996}) or WMAP and Planck flux measurements (absolute sensitivity of $\sim 7\times10^{-23}\,\mathrm{Wm}^{-2}\, \mathrm{Hz}^{-1}\, \mathrm{sr}^{-1}$, \citeauthor{Hinshaw2009} \citeyear{Hinshaw2009}; \citeauthor{Planck2013_Results} \citeyear{Planck2013_Results}). The total CMB distortion is about 0.2 and 1.7 $\mathrm{nWm}^{-2}\,\mathrm{sr}^{-1}$ for $z_{\mathrm{max}} = 6$ for Model A and B, respectively (Fig. 4). Model B predicts a faster increase of the total CMB distortion with $z_{\mathrm{max}}$ than Model A. The maximum distortion increases up to $z_{\mathrm{max}} \sim 7$. At higher $z$, the CMB is not distorted, because the model is effectively free of dust. Note that the reported values are the lower thresholds; the realistic distortions should be higher.

\section {Discussion}

It is commonly considered that the CMB is distorted by foreground diffuse FIR and submillimetre emission of dust in the Milky Way, other galaxies and intergalactic space \citep{Draine2009, Imara2016}. However, the CMB can also be distorted due to absorption by dust producing a decline of the CMB intensity at all frequencies. This distortion should be high enough to be observable in the CMB measurements. The maximum spectral distortion of the CMB light coming from $z = 10$ is predicted at 300 GHz being at least 20 times higher than the detection level of the COBE/FIRAS measurements \citep{Fixsen1996} and at least 35 times higher than the detection level of the WMAP or Planck measurements \citep{Hinshaw2009,Planck2013_Results}. The CMB should be distorted also in a perfectly transparent universe just due to absorption by dust in galaxies. This effect is about one order lower than that for the intergalactic opacity, but still above the detection level of the current CMB measurements.

Finally, let's shortly discuss why the imprint of dust is missing on the CMB. Firstly, we can speculate that the parameters used in modelling are seriously biased. However, it contradicts observations of the intergalactic opacity \citep{Menard2010a, Xie2015, Imara2015}, opacity of galaxies \citep{Gonzalez1998, Calzetti2001, Holwerda2005a, Holwerda2005b, Holwerda2013} and the extinction law data in the Milky Way \citep{Draine1984, Mathis1990, Li2001, Draine2003}. Secondly, we can question the Big Bang as the origin of the CMB and revive theory of the CMB as the thermal radiation of dust itself being produced at much later times than Big Bang \citep{Layzer1973, Wright1982, Wright1987, Wright1991, Aguirre2000}. In such theory, the CMB should not be distorted because the CMB would concurrently be absorbed and reradiated by dust. In any case, it is clear that the missing dust imprint on the CMB is an intriguing puzzle which should be further studied and confronted with  current measurements and models of the Universe.

\section*{Acknowledgements}
I thank Benne W. Holwerda for his valuable comments which helped me to improve the quality of the paper.


\bibliographystyle{mnras}


\end{document}